\documentclass[12pt,a4paper]{article}
\usepackage[utf8]{inputenc}
\usepackage[T1]{fontenc}
\usepackage[margin=1.9cm]{geometry}
\usepackage[pdfstartview={FitH},bookmarks=false,linktoc=page,colorlinks=true,linkcolor=blue,citecolor=blue,urlcolor=blue]{hyperref}
\usepackage[numbered]{bookmark}
\usepackage{mathtools}
\usepackage{amssymb}
\usepackage{graphicx}
\usepackage[font=small,labelfont=bf]{caption}
\usepackage{authblk}
\usepackage{cite}
\usepackage{dsfont}
\usepackage[dvipsnames]{xcolor} 
\usepackage[titles]{tocloft}
\usepackage{float}
\usepackage{subcaption}
\usepackage{color}
\usepackage{tikz}
\usepackage{ifthen}
\usepackage[warn]{textcomp}
\numberwithin{equation}{section}
\allowdisplaybreaks
\usepackage{multirow}
\usepackage{cancel}

\usetikzlibrary{patterns}

\newcommand{\dbar}{{\mkern3mu\mathchar'26\mkern-12mu d}}

\usepackage{comment}

\usepackage{caption}
\usepackage{subcaption}

\begin{document}
\title{\vspace{2cm}\textbf{ On Thermodynamic Stability of Black Holes. Part I: Classical Stability}\vspace{1cm}}

\author[a]{V. Avramov}
\author[a,b]{H. Dimov}
\author[a]{M. Radomirov}
\author[a,c]{R. C. Rashkov}
\author[a]{T. Vetsov}

\affil[a]{\textit{Department of Physics, Sofia University,}\authorcr\textit{5 J. Bourchier Blvd., 1164 Sofia, Bulgaria}
	
	\vspace{-10pt}\texttt{}\vspace{0.0cm}}

\affil[b]{\textit{The Bogoliubov Laboratory of Theoretical Physics, JINR,}\authorcr\textit{141980 Dubna,
		Moscow region, Russia}
	
	\vspace{-10pt}\texttt{}\vspace{0.0cm}}

\affil[c]{\textit{Institute for Theoretical Physics, Vienna University of Technology,}
  \authorcr\textit{Wiedner Hauptstr. 8–10, 1040 Vienna, Austria}
	\vspace{10pt}\texttt{v.avramov,h\_dimov,radomirov,rash,vetsov@phys.uni-sofia.bg}\vspace{0.1cm}}
\date{}
\maketitle

\begin{abstract}
We revisit the classical thermodynamic stability of the standard black hole solutions by  implementing the intrinsic necessary and sufficient conditions for stable global and local thermodynamic equilibrium. The criteria for such equilibria are quite generic and well-established in classical thermodynamics, but they have not been fully utilized in black hole physics. We show how  weaker or incomplete conditions could lead to misleading or incorrect results for the thermodynamic stability of the system. We also stress the importance of finding all possible local heat capacities in order to fully describe the classical  equilibrium picture of black holes. Finally, we thoroughly investigate the critical and phase transition curves and the limits of the classical analysis. This paper is the first in the line of intended works on thermodynamic stability of black holes in modified theories of gravity and holography.
\end{abstract}

\thispagestyle{empty}
\tableofcontents

\section{Introduction}
Black hole physics is definitely one of the major topics in modern theoretical physics. This is largely due to the fact that a consistent  description of their properties relies on fundamental principles from thermodynamics, statistical mechanics, quantum theory and gravitation. The interest in the area is also  significantly boosted by the recent  discovery of gravitational waves produced by black hole collisions \cite{LIGOScientific:2016aoc}, and further by the visual confirmation of the existence of supermassive black holes in the center of M87 and our own galaxy \cite{EventHorizonTelescope:2019dse, EventHorizonTelescope:2021dqv, EventHorizonTelescope:2022wkp}. Evidently, near-future experiments and observations are also expected to reveal surprising new results.

In this context thermodynamics of black holes \cite{Bekenstein:1972tm, Bekenstein:1973ur, Bekenstein:1974ax, Bekenstein:1975tw, Hawking:1971tu, Hawking:1975vcx, Hawking:1976de, Bardeen:1973gs, Davies:1977bgr, Davies_1978} could provide an avenue for direct tests of our theories of gravity.  The latter is a consequence of the generic feature of thermal systems to allow descriptions with only few observable parameters. For black holes these macro parameters reduce to the mass, entropy, charge, angular momentum and a few more variables depending on the underlying model of gravity. More importantly, the origins of black hole thermodynamic are deeply rooted in the theory of quantum gravity, which is  under active investigation by various promising approaches such as string theory. On the other hand, insights on the quantum nature of gravity can only become available near the event horizon, where strong gravitational effects become relevant and thermodynamics can no longer be treated classically. These arguments make black holes one of the most challenging objects to study.

As apparent by the images of supermassive black holes the thermodynamic states of such compact objects strongly depend on their surrounding environment. For example, this could be achieved by matter accretion onto the even horizon \cite{Luminet:1979nyg, Page:1974he, Thorne:1974ve, Gyulchev:2019tvk, Gyulchev:2021dvt}, or via some radiative process such as Hawking radiation \cite{Hawking:1975vcx}. In these cases, the natural question to ask is under what condition the system is in thermodynamic equilibrium with its surroundings? To our knowledge, the answer to this question has not been explored  in sufficient detail even for the standard black hole solutions, although the criteria for equilibrium are quite generic and well-established in standard thermodynamics \cite{bazarov1964thermodynamics, callen2006thermodynamics, greiner2012thermodynamics, swendsen2020introduction, blundell2010concepts}. 
There are only few known examples of using such generic criteria for black holes. For example, discussions on the necessary and sufficient conditions of thermodynamic stability for higher dimensional black holes have been presented in \cite{Dolan:2014lea, Dolan:2013yca}. A subsequent derivation of general stability criteria for black holes has been presented by A. K. Sinha \cite{Sinha:2015zxv, Sinha19aaa, Sinha21bb}. The latter coincide with the Sylvester criterion for positive definiteness of the Hessian of the energy potential. Applications to thermodynamic stability of quantum black holes appear in \cite{Sinha21bb,Sinha:2016evq, Sinha:2015zxv, Sinha:2017obh,
Sinha:2020uwf, Sinha19aaa}. Thermal stability of black holes with arbitrary hairs has been investigated by \cite{Sinha:2017cgi, Sinha:2017obh,Sinha:2020uwf}. Our work aims to supplement these studies by implementing the full classical criteria for local and global thermodynamic stability of the standard black hole solutions in General relativity. We also stress the importance of working in natural parameters for a given thermodynamic representation and raise some caution when working with non-generic or partial stability criteria.

In classical thermodynamics there are two types of equilibria -- local and global\footnote{We take on the definitions of ``local'' and ``global'' thermodynamic stability as presented by Callen \cite{callen2006thermodynamics}. He defines ``global'' condition of stability as the general convexity/concavity
of the energy/entropy (see Chapter 8 and Appendix \ref{appA}).}.
If a system resides in a global thermodynamic equilibrium then, by definition, it
has the same temperature, the same pressure, the same chemical potentials etc, everywhere
within its boundaries. In this case, one can study the global thermodynamic stability  in a given representation by considering the properties of the Hessians of the corresponding thermodynamic potentials. The global thermodynamic analysis is based on two equivalent criteria: the Hessian eigenvalue method and the Sylvester criterion for positive definiteness of quadratic forms. One can use both criteria independently as sufficient conditions for global thermodynamic stability of any systems including black holes.

The system is said to be in a local thermodynamic equilibrium if it can be divided into smaller
constituents, which are individually in approximate thermodynamic equilibrium. In each partial system the intensive thermodynamic
state quantities assume definite constant values and do not vary too strongly from one partial
system to another, i.e. only small gradients are allowed. The study of local thermodynamic stability is based on the admissible heat capacities. Specifically, for a system to be locally stable with respect to a perturbation in a set of parameters the corresponding heat capacities must be strictly positive.
 
It is important to note that local equilibrium does not imply a global one. On the other hand, it is natural to assume that a system in global thermodynamic equilibrium is also locally stable. This is evident by the fact that the components of the Hessians of the thermodynamic potentials can be related to the  local heat capacities of the system. 

The goal of this work is to present the theory of classical thermodynamic stability in details and then revisit the standard black hole solutions in the light of the necessary and sufficient condition for thermodynamic equilibrium\footnote{In \cite{Dimov:2021fbm} the authors used the weaker Sylvester criterion for semi-definite positive quadratic forms. In this paper we show that this could lead to some contradictions with the stability of the system, thus only the stronger criterion for positive definite forms should be taken into account.}. We show that all considered black hole solutions are globally unstable, but some local stability with respect to their heat capacity can be retained. The instability of asymptotically flat black holes is a persistent feature
in any dimension. For example in \cite{Dias:2010eu} for $D\geq 5$ it was shown that all asymptotically flat rotating and neutral black holes are unstable and this was extended
to include the charged case in \cite{Dolan:2014lea}. 

The structure of the paper is the following. In Section \ref{sec2} we present the necessary and sufficient conditions for global thermodynamic stability in energy and entropy representations\footnote{We follow Callen \cite{callen2006thermodynamics}, but the same criteria were derived from the partition function by A. K. Sinha \cite{Sinha:2015zxv, Sinha19aaa, Sinha21bb}.  }. In Section \ref{secSchw} we revisit the thermodynamic instability of the Schwarzschild black hole solution only as a didactic example. In Section \ref{secRNBH} we study the thermodynamic stability of the Reissner-Nordstr\"om (RN) black hole. We confirm its global instability, but show that its is locally stable with respect to certain processes. In Section \ref{secKerr} the thermodynamic stability of Kerr solution is studied in details. We verify that Kerr is globally unstable, but it has regions of local stability for particular values of the angular momentum. We also show that for processes with constant mass the $J\to 0$ is a regular limit to a new locally stable state, which differs from the Schwarzschild black hole. In Section \ref{secKN} we verify that even the three parametric thermodynamic space of equilibrium states for the Kerr-Newman (KN) solution is not enough to support global thermodynamic stability of the system. In this case we study all admissible heat capacities and derive the various regions of local thermodynamic stability, which have not been fully investigated previously. Finally, in Section \ref{secConcl} we give a brief summary of our results.

\section{Description of thermodynamic stability}\label{sec2}

In this section we present the necessary and
sufficient conditions for classical global and local thermodynamic stability in energy and entropy representations.

\subsection{Energy and entropy representations}

The thermodynamic  representation of a given system is defined by the choice of thermodynamic potential used to describe the properties of the system and the constraints it is subjected. Hence, the energy representation is used when the preferable thermodynamic potential is the internal energy $E$ of the system. In this case, one naturally imposes constraints  on the entropy and other extensive variables of the system. Consequent application of  Legendre transformation along one or several natural parameters of the internal energy leads to other energy derived thermodynamic representations, called free energies, which fully describe the properties of the system on their own. There exist other representations, which cannot be derived from the energy potential via Legendre transformation. Such representation is the entropy representation, where the entropy derived potentials are called Massieu-Planck potentials or free entropies. At the end the choice of a representation depends on the initial constraints imposed on the system. 

In energy representation one defines the set of all extensive\footnote{We assume the standard convention that vector-columns are vectors and vector-rows as their transpose.} $\vec E=(E^1,E^2,...,E^n)^T$ and all intensive $\vec I=(I_1,I_2,...,I_n)^T$ parameters, which describe the possible macro states of the system. In these terms the first law of thermodynamics in equilibrium is written by\footnote{For example, for an ideal gas, one has $dE=TdS-pdV+\mu dN$, hence $\vec I=(T,-p,\mu)^T$ and $\vec E=(S,V,N)^T$. }:
\begin{equation}\label{eqFLTDER}
dE= \sum\limits_{a=1}^n I_a dE^a=\vec I.d\vec E,
\end{equation}
where $E$ is the (internal) energy of the system.
The form of the first law is specifically chosen to represent $ I_a$ as  generalized thermodynamic forces and $ E^a$ as generalized thermodynamic coordinates by analogy of classical mechanics.
 When it is possible to express the energy potential $E$ as a function of its natural extensive variables $\vec E$, one finds the so called fundamental relation:
\begin{equation}\label{eqFundRelE}
E=E(E^1,E^2,...,E^n).
\end{equation}
This is a particularly important relation due to the fact that it can be used to directly extract the relevant thermodynamic properties of the system via the equations of state:
\begin{equation}\label{eqStateErep}
I_a=\frac{\partial E(\vec E)}{\partial E^a}\bigg|_{E^1,...,\hat E^a,...,E^n}.
\end{equation}
Here the parameters in the subscript are kept fixed except for $E^a$. The set of equations (\ref{eqFLTDER})-(\ref{eqStateErep}) define the mathematical form of the energy representation for a system in equilibrium.

However, it is not always practical to work with the energy, due to the fact that different constraints on the system may require different control parameters. In this case, it may be useful to transform to another representation without loosing any relevant thermodynamic information of the system. This can be archived by the well-known Legendre transformation. 
Performing Legendre transformation $\mathcal{L}$ along one or several natural parameters of the energy we can obtain all of the standard free energy potentials. In this case, the one-parameter Legendre family of energy derived potentials $\Phi_a$, is given by\footnote{We can again refer to the ideal gas thermodynamics, where $E=U$, $E^1=S$, $E^2=V$, hence $\Phi_1\equiv F=\mathcal{L}_S U=U-T S$ is the Helmholtz free energy, and  $\Phi_2\equiv H=\mathcal{L}_V U=U+(-p) V=U+p V$ is the enthalpy.}:
\begin{align}
&\Phi_1=\mathcal{L}_{E^1}E=E-I_1 E^1,\\
 &\Phi_2=\mathcal{L}_{E^2}E=E-I_2 E^2,\\\nonumber
 &\vdots
 \\
 &\Phi_n=\mathcal{L}_{E^n}E=E-I_n E^n.
\end{align}
Consequently, the two-parameter Legendre family of energy derived potentials $\Phi_{ab}$, $a\neq b$, is\footnote{To make the analogy full one refers to $\Phi_{1,2}\equiv G=\mathcal{L}_{S,V} U=U-T S+pV=F+pV=H-TS$ as the Gibbs free energy, and $\Phi_{1,3}\equiv \Omega=\mathcal{L}_{S,N} U=U-T S-\mu N=F-\mu N$ as the grand potential.}:
\begin{align}
	&\Phi_{1,2}=\mathcal{L}_{E^1,E^2}E=E-I_1 E^1-I_2 E^2,
	\\
	&\Phi_{1,3}=\mathcal{L}_{E^1,E^3}E=E-I_1 E^1-I_3 E^3, \\\nonumber
	\vdots
	\\
	&\Phi_{n-1,n}=\mathcal{L}_{E^{n-1},E^{ n}}E=E-I_{n-1} E^{n-1}-I_n E^n.
\end{align}
This approach can be generalized to any number of extensive variables. At the end, the Legendre transformation along all the extensive variables leads to the trivial (or null) potential  $\Phi_{1,2,...,n}=0$, which is due to the Euler homogeneity relation
\begin{equation}
	E=\sum\limits_{a=1}^n I_a E^a.
\end{equation}
However, the potential $\Phi_{1,2,...,n}$ may not be trivial if the energy of the system is a quasi-homogeneous function of degree $r$ and type $(r_1,...,r_n)$, i.e.  
\begin{equation}
    E(\tau^{r_1} E^1,\tau^{r_2} E^2,...,\tau^{r_n} E^n)=\tau^r E(E^1,E^2,...,E^n),
\end{equation}
where under dilatations by a scale factor $\tau>0$ one has the generalized Euler relation:
\begin{equation}
r E=\sum\limits_{a=1}^n r_a I_a E^a.
\end{equation}
In black hole thermodynamics this is related to the so called Smarr relation\footnote{See for example \cite{Dolan:2014jva, Dolan:2014woa}.}. 

The situation is similar if one works in the entropy representation. In this case, choosing the entropy $S$ as a thermodynamic potential depending on its natural extensive parameters $S=S(S^1,S^2,...,S^n)$, one can write the first law of thermodynamics in the form
\begin{equation}\label{eqFirstLawS}
	d S=\sum\limits_{a=1}^n \lambda_a dS^a=\vec \lambda . d\vec S,
\end{equation}
where the intensive variables $\vec \lambda=(\lambda_1,\lambda_2,...,\lambda_n)^T$ are the thermodynamically conjugate parameters of $\vec S$. The equations of state follow naturally by\footnote{Note that $\lambda_1=\dfrac{1}{T},\, \lambda_b=-\dfrac{I_b}{T}$, where $b=2,3,...,n$.}
\begin{equation}
	\lambda_a=\frac{\partial S(\vec S)}{\partial S^a}\bigg|_{S_1,...,\hat S^a,...,S^n}.
\end{equation}

Legendre transformation of the entropy potential along one or several of its natural parameters is used to obtain the entropy derived family of potentials. The latter are known by several names: Massieu-Planck potentials, free entropies or free information potentials. It is important to note that entropy is not a Legendre transformation of the energy and thus entropy representation and its derived potentials are generally different from the energy representation related potentials. In fact, different potentials correspond to different constraints to which
the system may be subjected\footnote{Thermodynamic potentials are naturally used to describe the ability of a system to perform some kind of
work under given constraints. These constraints are usually the constancy of some state variables like pressure,
volume, temperature, entropy, etc. Under such conditions the decrease in
thermodynamic potential  from one state to another is equal to the amount of work that is produced when a
reversible process carries out the transition, and hence is the upper bound to the amount of work produced by
any other process, \cite{Andresen1999}.}. The thermodynamic properties of the system can be fully described once the fundamental relation in the  chosen representation  has been established. 

Let us clarify this point with a simple example. Assume that we want to study Kerr black hole with first law in energy representation  given by
\begin{equation}
    dM=T dS+\Omega dJ.
\end{equation}
In this case, the natural parameters of the mass are the entropy $S$ and the angular momentum $J$. Hence the equilibrium manifold is defined by the embedding of the fundamental relation  $M=M(S,J)$, which is a two-dimensional surface in $\mathbb{R}^3$. This representation is useful to study the stability and the critical phenomena of the system with respect to $S$ and $J$. This means that it is not a good idea to look for the critical properties of the temperature in this representation, because $T$ is not a natural variable of the mass\footnote{It is well known that one may loose information of the system if not working in natural variables.}. In such cases one looks for an appropriate thermodynamic potential, whose natural variable is $T$. For example, one can Legendre transform the mass to the Helmholtz free energy (canonical ensemble), $F=\mathcal{L}_S M=M-T S$, with first law
\begin{equation}
    dF=-S dT+\Omega dJ.
\end{equation}
It is now evident that the natural space is $F=F(T,J)$ and one can use $F$ to study the properties of the system in terms of the temperature. Similarly, one can refer to the Gibbs free energy, etc.

We are now ready to describe the generic conditions for thermodynamic stability.

\subsection{Classical criteria for global thermodynamic stability}

We say that a thermodynamic system is in equilibrium with its surroundings if the state quantities do not spontaneously change over considerably long period of time.
According to the laws of thermodynamics \cite{bazarov1964thermodynamics, callen2006thermodynamics, greiner2012thermodynamics, swendsen2020introduction, blundell2010concepts} the necessary, but not sufficient, conditions for thermodynamic equilibrium between the system and its surroundings can be established by the equalities of the corresponding intensive parameters, $I_a=I_a^*$, of the system $I_a$ and the reservoir $I_a^*$. These parameters may include temperature, pressure, chemical potentials etc. The conditions can easily be derived by the restriction on the first variation of the internal energy of the system during a virtual process:
\begin{equation}
	\delta^{(1)} E(E^a)-\sum_aI_a^* \delta E^a=\sum_a\bigg[\bigg(\frac{\partial E}{\partial E^a}\bigg|_{E^1,...,\hat E^a,...,E^n}-I_a^*\bigg) \delta E^a \bigg]=0.
\end{equation}

The space of possible states of equilibrium (compatible with constraints and initial conditions) is called
the space of virtual states. Due to the first law in equilibrium one has (\ref{eqStateErep}), thus the necessary conditions for equilibrium become
\begin{equation}
	I_a=I_a^*=const.
\end{equation}
One can reach to the same  conclusion in the entropy representation by $\delta^{(1)} S(\vec S)=0$. 

On the other hand, the sufficient conditions for global thermodynamic equilibrium, and thus global thermodynamic stability, can be derived by the sign of the second variation of the energy or the entropy consistent with the second law of thermodynamics. Considering the energy as a potential the second variation
\begin{equation}
    \delta^{(2)}E=\delta \vec E^T.\mathcal{H}^{E}(\vec E).\delta \vec E>0 
\end{equation}
should be strictly positive due to the fact that in equilibrium the energy of the system assumes its minimum. Here $\mathcal{H}^E$ is the symmetric $n\times n$ Hessian matrix of the energy given by
\begin{equation}
\mathcal{H}^E_{ab}(\vec E)=\frac{\partial^2 E(\vec E)}{\partial E^a \partial E^b}\bigg|_{E^1,...,\hat E^a,...,\hat E^b,...,E^n},\quad a,b=1,2,...,n.
\end{equation}
The inequality $\delta^{(2)}E>0$ defines $\mathcal{H}^{E}$ as a positive definite quadratic form. This means that for global equilibrium it is  sufficient that all eigenvalues\footnote{Positive definiteness is sufficient but not necessary for the energy to be strictly convex.} $\varepsilon_a>0$, $a=1,...,n$,  of the Hessian of the energy be strictly positive. 

In the entropy representation the second variation 
\begin{equation}
\delta^{(2)}S=\delta \vec S^{\,T}.\mathcal{H}^{S}(\vec S).\delta \vec S<0
\end{equation}
should be strictly negative due to the fact that in equilibrium the entropy of the system settles at its maximum. The inequality $\delta^{(2)}S<0$ defines $\delta^{(2)}S$ as a negative definite quadratic form. For establishing a global equilibrium it is  sufficient that all eigenvalues $s_a<0$, $a=1,...,n$,  of the Hessian of the entropy be strictly negative.

An alternative set of  sufficient conditions for global\footnote{The reason why we call this criterion global has been explained in Appendix \ref{appA}.} thermodynamic stability is given by the Sylvester criterion for positive/negative definiteness of the Hessians. In energy representation the energy defines a global convex function, thus the Hessian of the energy is positive definite quadratic form. Therefore, all of  the principal minors $\Delta_k>0$ of the Hessian of the energy must be strictly positive. In entropy representation this criterion has alternating signs $(-1)^k \Delta_k>0$ due to the fact that entropy is globally concave function\footnote{The robustness of the stability criteria make them suitable to study any thermal system with well-defined first law of thermodynamics.}. An example is shown in Appendix \ref{appA}.

\subsection{Heat capacities and local thermodynamic stability}

One of the major effects of heat transfer is temperature change defined by $\dbar Q=C dT=T dS$, where the extensive quantity $C$ is called the total heat capacity of the system.  One can also write
\begin{equation}\label{eqTotalCapacity}
    C=\frac{\dbar Q}{dT}=T\frac{\partial S}{\partial T},
\end{equation}
and since $\dbar Q$ depends on the nature of the process\footnote{Hence the inexact differential $\dbar$.}, so does $C$. Hence, for different processes one has different heat capacities. The general definition of a heat capacity $C_{x^1,x^2,...,x^{n-1}}$, at fixed set of thermodynamic parameters $(x^1,x^2,...,x^{n-1})$, is given by the derivative of the entropy in a certain space of variables $(y^1,y^2,...,y^{n})$, namely\footnote{It is not necessary for the entropy to be a function of the variables $y^i$. The Nambu brackets automatically account for the Jacobians of the coordinate transformations from some other coordinates to the $y^i$ space. } \cite{Mansoori:2014oia}:
\begin{equation}
	C_{x^1,x^2,...,x^{n-1}}(y^1,y^2,...,y^{n})=T \frac{\partial S}{\partial T}\bigg|_{x^1,x^2,...,x^{n-1}}=T \frac{\{S,x^1,x^2,...,x^{n-1}\}_{y^1,y^2,...,y^n}}{\{T,x^1,x^2,...,x^{n-1}\}_{y^1,y^2,...,y^n}},
\end{equation}
The Nambu brackets \{ \}, used in the formula above, generalize the Poisson brackets for three or more independent variables (see Appendix \ref{appB}). Furthermore, the set of  constant parameters $x^1,x^2,...,x^{n-1}$ could be a mix of all kinds of intensive and extensive variables. Additionally, all the relevant state quantities become functions of the independent  parameters $y^1,y^2,...,y^{n}$. In this case we say that $(y^1,...,y^n)$ define the coordinates of our space of equilibrium states. 

Local thermodynamic equilibrium can be defined by quasi equilibrium between different parts of the system, where sufficiently small gradients of the parameters are still allowed. The identification of local stability with the positivity of certain heat capacity is related to the components of the Hessian, where imposing the generic conditions for stability always require $C>0$. This is most evident for simple systems (see for example \cite{bazarov1964thermodynamics, callen2006thermodynamics, greiner2012thermodynamics, swendsen2020introduction, blundell2010concepts}). Therefore, one can insist that the classical condition for local thermodynamic stability, with respect to some fixed parameters ($x^1,x^2,...,x^{n-1}$), is\footnote{Note that even if all admissible heat capacities are positive in a given range of parameters, global equilibrium may still be absent.}
\begin{equation}
C_{x^1,x^2,...,x^{n-1}}>0.
\end{equation}

Heat capacities are also important for identifying critical and phase structures in the system. Specifically, if a given heat capacity diverges\footnote{Paul Davies 
first pointed out that the divergence of the the heat capacity of the Kerr-Newman black
hole is a mark of a second order phase transition \cite{Davies:1977bgr}.} or changes sign this would signal the presence of a phase transition and the breakdown of the equilibrium thermodynamic description.

In the forthcoming sections we are going to revisit the thermodynamic stability  of the standard black hole systems from general relativity in the light of the strict classical criteria presented above.
 
\section{Thermodynamic instability of the Schwarzschild solution}\label{secSchw}

The simplest space-time solution of general theory of relativity is the Schwarzschild solution
\begin{equation}
    ds^2=-\bigg(1-\frac{2M}{r}\bigg)dt^2+\bigg(1-\frac{2M}{r}\bigg)^{-1}dr^2+r^2 (d\theta^2+\sin^2\theta d\varphi^2),
\end{equation}
which describes a static spherically symmetric black hole. It is purely didactic to study its thermodynamic properties, which are defined only by three parameters: the mass $M$, the entropy $S$ and the Hawking temperature\footnote{The third law of thermodynamics insists on $T>0$.} $T$ on the event horizon of the black hole. In the energy representation the mass of the black hole is identified by the energy of the system, hence in equilibrium  the first law of thermodynamics is simply written by\footnote{In the extended thermodynamics, where the cosmological parameter is treated as pressure, the mass of the black hole is identified by the enthalpy of spacetime \cite{Cvetic:2010jb}.}
\begin{equation}
    dM= TdS.
\end{equation}
In terms of the energy natural parameter $S$ one has
\begin{align}\label{eqSTSchw}
M=\frac{\sqrt{S}}{2\sqrt{\pi}}, \quad T=\frac{\partial M}{\partial S}= \frac{1}{4\sqrt{\pi S}}.
\end{align}
%

Assuming $M, S,T>0$ the Schwarzschild thermodynamics can be presented in a more convenient form
\begin{equation}
m=\sqrt{s}, \quad \tau=\frac{\partial m}{\partial s}=\frac{1}{2\sqrt{s}},\quad dm=\tau ds,
\end{equation}
where we have introduced the following notations:
\begin{equation}\label{RescaledShwarz}
m=2M, \quad \tau=2\pi T, \quad  s=\frac{S}{\pi}.
\end{equation}

The global thermodynamic instability of the Schwarzschild black hole solution follows directly from the Hessian of the mass, which has only one element
\begin{equation}
\mathcal H_{ss}=\frac{\partial^2m }{\partial s^2}=-\frac{1}{4s^{3/2}}<0.
\end{equation}
It is evident that the sign contradicts the general thermodynamic stability criteria (\ref{eqGlobalCondE})-(\ref{eqGlobalCondE1}). 

In order to analyze the local thermodynamic stability one looks at the heat capacity of the system
\begin{equation}\label{HeatCapSch}
C=\tau\frac{\partial s}{\partial \tau}= \tau \bigg( \frac{\partial\tau}{\partial s} \bigg)^{-1}=-2s<0.
\end{equation}
Its negative sign defines the local thermodynamic instability of the Schwarzschild black hole, so it can radiate. One can think of this as a process of evaporation where the temperature $T$ rises on the expense of the decreasing mass $M$ of the black hole\footnote{Hence the negative sign of the gradient $\partial M/\partial T<0$.}. At the end the system will evaporate explosively unless quantum effects are taken into account. The
Schwarzschild black hole can be stabilized by placing it in a cavity with a heat
bath at a finite distance from the horizon \cite{York:1986it} or in the presence of negative
cosmological constant \cite{Hawking:1982dh} of sufficient magnitude. Quantum corrections also lead to thermodynamically stable Schwarzschild configurations \cite{El-Menoufi:2017kew, Calmet:2021lny, Xiao:2021zly, Berry:2021hos}.

Next we revisit the classical thermodynamic stability of the Reissner-Nordstr\"om and the Kerr black hole solutions.

\section{Thermodynamic stability of Reissner-Nordstr\"om solution}\label{secRNBH}

\subsection{Thermodynamics in Reissner-Nordstr\"om spacetime}

Reissner-Nordstr\"om (RN) solution is the charged generalization of the Schwarzschild black hole\footnote{The metric is written in spherical coordinates.}:
\begin{equation}
    ds^2=-\bigg(1-\frac{2M}{r}+\frac{Q^2}{r^2}\bigg)dt^2 +\bigg(1-\frac{2M}{r}+\frac{Q^2}{r^2}\bigg)^{\!-1}dr^2 +r^2 (d\theta^2+\sin^2\theta d\varphi^2).
\end{equation}
Here $M>0$ is the mass and $Q\in \mathbb R$ is the charge of the black hole. The event horizon is located at $r_+=M+\sqrt{M^2-Q^2}$ and its  existence ($r_+>0$) assumes the condition $M^2>Q^2$. 
Note that the extremal case $M=|Q|$ leads to $T=0$, which is in contradiction to the third law of thermodynamics\footnote{For a discussion of how the third law can be violated by black holes see \cite{Davies_1978}. Extremal cases might still be interesting for string theory and other approaches to quantum gravity.}. In equilibrium the first law yields
\begin{equation}\label{eqFLRNM}
    dM=T dS+\Phi dQ,
\end{equation}
where $S$ is the entropy, $T$ is the Hawking temperature, and $\Phi$ is the electric potential of the black hole. The relation (\ref{eqFLRNM}) defines the thermodynamics of the RN black hole in ($S,Q$) space:
\begin{align}
M= \frac{S +\pi Q^2}{2 \sqrt{\pi S }},
\quad T=\frac{\partial M}{\partial S}\bigg|_Q=\frac{S-\pi  Q^2}{4 \sqrt{\pi } S^{3/2}}, \quad 
\Phi=\frac{\partial M}{\partial Q}\bigg|_S=\frac{\sqrt{\pi } Q}{\sqrt{S}}.
\end{align}

\begin{figure}[H]
\centering
\includegraphics[scale=0.34]{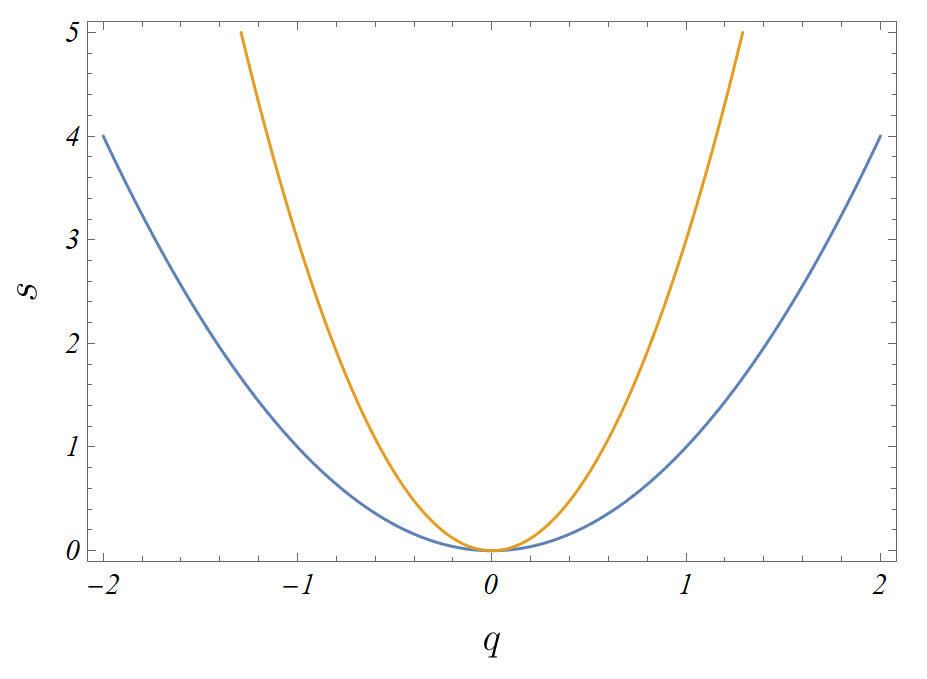}
\caption{ The existence of the black hole is defined above the blue parabola $s=q^2$. The orange parabola $s=3 q^2$ represents the Davies curve for the heat capacity $C_q$. One has local thermodynamic stability with respect to fixed charge $q=const$ in the region between the blue and the orange curves. For fixed mass $m=const$ RN is locally stable everywhere above the blue curve. No locally stable regions exist for RN with respect to constant electric potential $\phi$.   }
\label{RNBlackHole}
\end{figure}

After introducing the set of new parameters,
\begin{equation}\label{RescaledParRN}
m=2M, \quad \tau=2\pi T, \quad  s=\frac{S}{\pi}, \quad \phi=2\Phi, \quad q=Q,
\end{equation}
one can write the first law in the form $dm=\tau ds +\phi dq$, where
\begin{align}\label{eqMassRN}
&m= \frac{s +q^2}{ \sqrt{s} },
\\ \label{eqTinSQRN}
&\tau =\frac{\partial m}{\partial s}\bigg|_q=\frac{s -q^2}{2s^{3/2}},
\\
\label{eqFinSQRN}
&\phi=\frac{\partial m}{\partial q}\bigg|_s=\frac{ 2q}{\sqrt{s}}.
\end{align}
In this representation the existence condition $M^2>Q^2$ becomes
\begin{equation}\label{eqRegExistRNBH}
    s>q^2.
\end{equation}
The latter is satisfied above the blue parabola in  $(s,q)$ space (Fig.\!\! \ref{RNBlackHole}). Note that the curve $s=q^2$ is forbidden by the third law of thermodynamics. The condition for existence (\ref{eqRegExistRNBH}) together with (\ref{eqMassRN}) and (\ref{eqFinSQRN}) also lead to  
\begin{equation}
\sqrt{s}<m<2\sqrt{s}, \quad  m>2|q|, \quad |\phi|<2.
\end{equation}

\subsection{Global thermodynamic instability of Reissner-Nordstr\"om black hole} \label{RNsect}

The global thermodynamic stability of the RN black hole solution can be determined by the properties of the  Hessian of the mass in ($s,q$) space:
\begin{align}
\mathcal{H}(s,q)&=\left(\! \begin{array}{cc}
\mathcal{H}_{ss} & \mathcal{H}_{sq}\\[5pt]
\mathcal{H}_{qs} & \mathcal{H}_{qq}\\
\end{array} \!\right)=
\left(\!
\begin{array}{ccc}
\frac{\partial^2 m}{\partial s^2} & \frac{\partial^2 m}{\partial s \partial q}\\[5pt]
\frac{\partial^2 m}{\partial q\partial s} & \frac{\partial^2 m}{\partial q^2}
\end{array}
\!\right)=\left(\!
\begin{array}{cc}
 \frac{3q^2 -s}{4 s^{5/2}} & -\frac{q}{ s^{3/2}} \\[5pt]
 -\frac{q}{ s^{3/2}} & \frac{2}{\sqrt{s}} \\
\end{array}
\!\right).
\end{align}
According to the general theory the sufficient conditions for having a stable global equilibrium in the mass-energy representation insists on $\lambda_1>0$ and $\lambda_2>0$. The two eigenvalues of $\mathcal{H}$ are
\begin{align}
    \lambda_{1,2}=\frac{3 q^2-s (1-8 s)\pm\sqrt{9 q^4+2 q^2 s (8 s-3)+s^2 (8 s+1)^2}}{8 s^{5/2}}.
\end{align}
A closer inspection of their signs shows that $\lambda_1>0$ and $\lambda_2<0$ for all admissible values of $s$ and $q$ in the region of existence (\ref{eqRegExistRNBH}) of the black hole. This indicates  that the Reissner-Nordstr\"om black hole cannot be globally stable from thermodynamic standpoint. 

An additional check of the Sylvester criterion for the positive definiteness of $\mathcal{H}$ also confirms our this. The conditions for global thermodynamic stability require:
\begin{equation}\label{eqFullSylCRN}
    \mathcal{H}_{ss} = \frac{3q^2-s}{4s^{5/2}}> 0,\quad  \mathcal{H}_{qq}=\frac{2}{\sqrt{s}}> 0,\quad \det\mathcal{H}=\frac{q^2-s}{2s^3}> 0.
\end{equation}
In this case, the first condition $\mathcal H_{ss}>0$ is satisfied below the orange parabola $s<3q^2$ (Fig. \ref{RNBlackHole}).
The second condition $\mathcal H_{qq}>0$ is always true. The final condition can not be true in the region of existence  \eqref{eqRegExistRNBH}, which is evident from the determinant of the Hessian:
\begin{equation}
    \det\mathcal{H}=-\frac{s -q^2}{2s^3}<0 \quad\text{for}\quad  s>q^2.
\end{equation}
Hence, global thermodynamic stability criteria (\ref{eqGlobalCondE})-(\ref{eqGlobalCondE1}) cannot be simultaneously satisfied and the RN solution is globally unstable from thermodynamic point of view. 

A note of caution is advised here when using the Sylvester criterion. It would be misleading to consider only the weak convexity conditions of the Hessian of the mass along $s$ and $q$: 
\begin{equation}\label{eqWeakTDSconds}
   \mathcal{H}_{ss}> 0,\quad  \mathcal{H}_{qq}> 0,
\end{equation}
which are satisfied in the region $q^2<s< 3q^2$. This would falsely indicate that RN is globally stable for $q^2<s< 3q^2$. Furthermore, it is not recommended to weaken the strict positive definiteness of the Hessian by positive semi-definiteness. The latter may incorrectly indicate that the system is stable on some of the critical or phase transition curves.

\subsection{Local thermodynamic stability of Reissner-Nordstr\"om black hole}

The local thermodynamic stability of the RN black hole is determined by the admissible heat capacities of the solution in $(s,q)$ space. By definition, for a fixed parameter $x$, one has
\begin{equation}
C_{x }(s,q)=\tau \frac{\partial s}{\partial \tau}\bigg|_x =\tau \frac{ \{s,x\}_{s,q}}{\{\tau,x \}_{s,q}},
\end{equation}
where the Nambu brackets are given by simple Poison brackets (Jacobians):
\begin{align}
&\{s,x\}_{s,q}=   \left|
\begin{array}{cc}
\frac{\partial s}{\partial s} & \frac{\partial s}{\partial q} \\[5pt]
\frac{\partial x}{\partial s} & \frac{\partial x}{\partial q} \end{array}
\right| =\left|
\begin{array}{cc}
1& 0\\[5pt]
\frac{\partial x}{\partial s} & \frac{\partial x}{\partial q} \end{array}
\right|, \quad  \{\tau,x\}_{s,q}=   \left|
\begin{array}{cc}
\frac{\partial \tau}{\partial s} & \frac{\partial \tau}{\partial q}\\[5pt]
\frac{\partial x}{\partial s} & \frac{\partial x}{\partial q} \end{array}
\right|.
\end{align}
Therefore, the relevant heat capacities  of the RN black hole in $(s, q)$ space are\footnote{Note that there are two more heat capacities, namely $C_s=0$ and $C_\tau=\infty$.}:
\begin{align}
\label{eqHCRN2}
 C_m &=\tau \frac{\partial s}{\partial \tau}\bigg|_{m}\!=\tau \frac{\{s,m\}_{s,q}}{\{\tau,m\}_{s,q}}=\frac{s \left(s - q^2\right)}{q^2},\\
\label{eqHCRN1}
 C_\phi&=\tau \frac{\partial s}{\partial \tau}\bigg|_{\phi }=\tau \frac{\{s,\phi\}_{s,q}}{\{\tau,\phi\}_{s,q}}= -2s,\\
\label{eqHCRN3}
C_q &=\tau \frac{\partial s}{\partial \tau}\bigg|_{q}=\tau \frac{\{s,q\}_{s,q}}{\{\tau,q\}_{s,q}} =\frac{2s ( s -q^2)}{3q^2 -s}.
\end{align}

For a processes with constant mass one notes that $C_m$ is always positive in the region of existence \eqref{eqRegExistRNBH}. Thus RN is locally stable above the blue parabola (Fig. \ref{RNBlackHole}).  Using the relation \eqref{eqMassRN} we can express $C_m$ as a function of only one variable $s$ or $q$ and the constant parameter $m_c$, i.e.
\begin{equation}
C_m= \frac{s \big(2 \sqrt{s} -m_c \big)}{ m_c -\sqrt{s}} =\frac{\big( m_c^2 -2q^2 +m_c\sqrt{m_c^2-4q^2}\, \big) \big( m_c^2 -4q^2 +m_c\sqrt{m_c^2-4 q^2}\, \big)}{4 q^2}.
\end{equation}

For a processes with constant electric potential the heat capacity $C_{\phi}$ is always negative, which leads to unstable black hole for this kind of processes. Using \eqref{eqFinSQRN} we can find $C_{\phi}$ as a function of the electric charge $q$ and the constant parameter $\phi_c$
\begin{equation}
C_\phi= -\frac{8q^2}{\phi_c^2}.
\end{equation}

Finally, when the charge is fixed, the region of local stability $C_q>0$ is $q^2<s<3q^2$.
It is located between the blue and the orange parabolas (Fig. \ref{RNBlackHole}). Above the orange parabola the RN black hole is unstable for processes with constant charge.

\subsection{Critical curves and phase transitions of Reissner-Nordstr\"om black hole}

The critical curves, also known as Davies curves, are defined by the divergences of the heat capacities ($C\to \pm \infty$) or by the curves where a change of sign occurs ($C=0$). In the $(s,q)$ space the RN black hole has the following critical curves:
\begin{align}
    C_m\to\left\{
\begin{array}{l}
     0,\,\,\,s \to q^2,
     \\
     \!\!\infty, \,\,q\to 0,
\end{array} 
\right.\quad
C_{\phi}\to \{0, \,\,q\to 0,
\quad
C_q\to\left\{
\begin{array}{l}
\,\,\,0,\quad s\to q^2,\\
\!\!\pm\infty, \,\,\,s\to 3q^2 \mp 0. 
\end{array} 
\right.
\end{align}

The heat capacities $C_m$ and $C_q$ change sign on the blue parabola $s=q^2$ (Fig. \ref{RNBlackHole}), where the temperature of the black hole is zero. The latter corresponds to the extremal case, which cannot be reached for
finite number of fluctuations due to the third law of thermodynamics. If we consider a process with constant mass, the line $q= 0$ is a Davies curve. On this line the RN black hole admits a phase transition, where the classical equilibrium description breaks down.

For processes with constant electric potential \eqref{eqFinSQRN} the limit $q\to0$ leads to $s\to 0$, which is forbidden by the third law of thermodynamics.

For processes with constant electric charge the orange parabola $s=3q^2$ is a Davies curve. Here RN phases from locally stable (below the orange curve) to locally unstable (above the orange curve) thermodynamic state.

\section{Thermodynamic stability of Kerr solution}\label{secKerr}

\subsection{Thermodynamics in Kerr spacetime}

The Kerr solution describes a rotating uncharged axially symmetric black hole with metric\footnote{In Boyer-Lindquist coordinates.}
\begin{equation}
    ds^2=-\bigg(1-\frac{2 M r}{\Sigma}\bigg) dt^2-\frac{4 M r a \sin^2\theta}{\Sigma} dt d\phi+\frac{\Sigma}{\Delta} dr^2+\Sigma d\theta^2+\frac{A \sin^2\theta}{\Sigma} d\phi^2.
\end{equation}
Here we use the standard notations: 
\begin{equation}
    \Sigma=r^2+a^2 \cos^2\theta,\quad \Delta=r^2-2 M r+a^2, \quad A=(r^2+a^2)^2-a^2\Delta \sin^2\theta,\quad J=a M.
\end{equation}

The existence of the event horizon, $r_+=M+\sqrt{M^2-a^2}$, leads to $M>\sqrt{|J|}$.
We do not include the extremal case $M= \sqrt{|J|}$ ($a=M$) due to the violation of the third law of thermodynamics. In the energy representation the first law is written by
\begin{equation}
    dM=T dS+\Omega dJ,
\end{equation}
where the parameters of the solution in ($S,J$) space take the form:
\begin{align}\label{eqKerrPar}
& M\!= \!\sqrt{\frac{4 \pi ^2 J^2+S^2}{4 \pi S\, }}, \,\,\,T\!=\!\frac{\partial M}{\partial S}\bigg|_J\!=\!\frac{S^2-4 \pi ^2 J^2}{4\sqrt{\pi S^3(4 \pi ^2 J^2+S^2)}}, \,\,\,\Omega\!=\!\frac{\partial M}{\partial J}\bigg|_S\!=\!\frac{2 \pi ^{3/2} J}{ \sqrt{S(4 \pi ^2 J^2+S^2)}}.
\end{align}

\begin{figure}[H]
\centering
\includegraphics[scale=0.35]{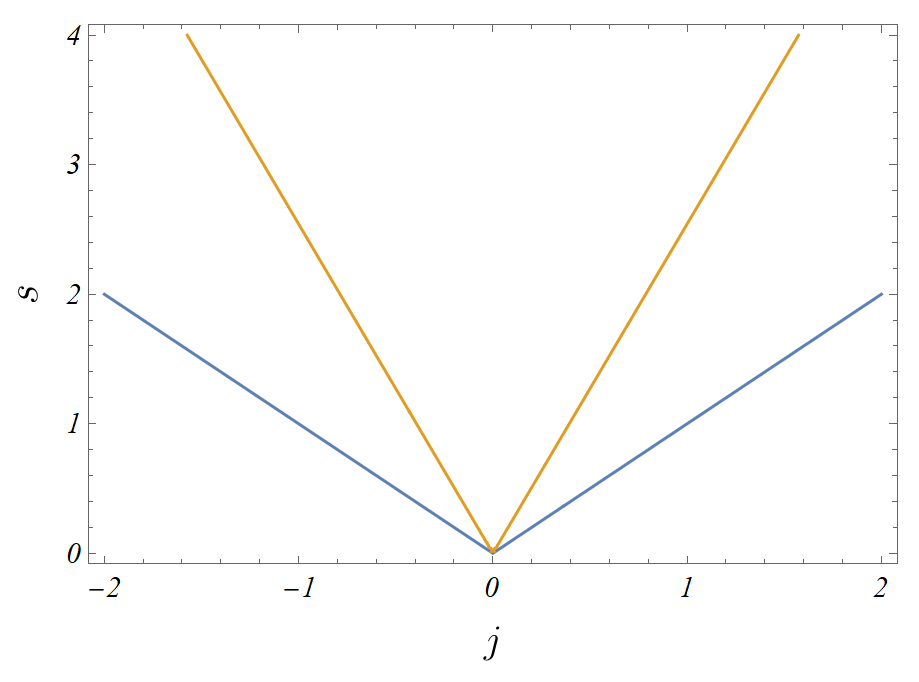}
\caption{ The existence of the black hole is defined above the blue lines  $s=|j|$. The orange lines $s=$ $|j|\sqrt{3+2 \sqrt{3}}$ represent the Davies curves for $C_j$. One has local thermodynamic stability with respect to fixed angular momentum $j=const$ in the region between the blue and the orange lines. For fixed mass $m=const$ Kerr is locally stable everywhere above the blue lines. No locally stable regions exist for Kerr with respect to constant angular velocity $\omega$.  }
\label{KerrBlackHole}
\end{figure}

To simplify the expressions we introduce new parameters:
\begin{equation}\label{RescaledParK}
m=2M, \quad \tau=2\pi T, \quad  s=\frac{S}{\pi}, \quad \omega=\Omega, \quad j=2J,
\end{equation}
where $ dm=\tau ds +\omega dj$ and
\begin{align}\label{eqMKerr}
&m= \sqrt{\frac{j^2+s^2}{s} },
\\ \label{eqTKerr}
&\tau =\frac{\partial m}{\partial s}\bigg|_j=\frac{s^2 -j^2}{2\sqrt{s^3 (s^2 +j^2)}},
\\
\label{eqOKerr}
&\omega=\frac{\partial m}{\partial j}\bigg|_s=\frac{j}{\sqrt{s(s^2 +j^2)}}.
\end{align}
Insisting on $s,\tau>0$ the existence of the Kerr black hole in in $(s,j)$ space is determined by
\begin{equation}
\label{eqRegExistKerr}
    s>|j|,
\end{equation}
which is satisfied above the blue line depicted on Fig. \ref{KerrBlackHole}. 
The condition for existence \eqref{eqRegExistKerr}
together with \eqref{eqMKerr} and \eqref{eqOKerr} also leads to
\begin{equation}
\sqrt{s}<m<\sqrt{2s}, \quad m>\sqrt{2|j|}, \quad |\omega|< \frac{1}{\sqrt{2s}}, \quad  |\omega|< \frac{1}{\sqrt{2|j|}}.
\end{equation}

\subsection{Global thermodynamic instability of Kerr black hole}

The Hessian of the mass in $(s,j)$ space is given by
\begin{equation}
\mathcal{H}=\left(\!\!\!\begin{array}{cc}
\mathcal{H}_{ss} & \mathcal{H}_{sj}\\[5pt]
\mathcal{H}_{js} & \mathcal{H}_{jj}\\
\end{array}\!\!\!\right)=
\left(\!\!
\begin{array}{ccc}
\frac{\partial^2 m}{\partial s^2} & \frac{\partial^2 m}{\partial s \partial j}\\[5pt]
\frac{\partial^2 m}{\partial j \partial s} & \frac{\partial^2 m}{\partial j^2} 
\end{array}
\!\!\right)
=\left(\!\!\!
\begin{array}{cc}
 \frac{3j^4 +6j^2s^2-s^4}{4\sqrt{s^5(s^2 +j^2)^3}} & -\frac{j(3s^2 +j^2)}{2\sqrt{s^3(s^2 +j^2)^3}} \\[7 pt]
 -\frac{j(3s^2 +j^2)}{2\sqrt{s^3(s^2 +j^2)^3}} &  \frac{\sqrt{s^3}}{\sqrt{(s^2 +j^2)^3}}  \\
\end{array}
\!\!\!\right).
\end{equation}
The two eigenvalues of the Hessian
\begin{equation}
    \lambda_{1,2}=\frac{3 \sqrt{j^2+s^2}\pm\sqrt{9 j^2+25 s^2}}{8 s^{5/2}}
\end{equation}
differ by signs: $\lambda_1>0$ and $\lambda_2<0$, for all admissible values of $s$ and $j$. This is sufficient to indicate that Kerr black hole is globally unstable solution. The same conclusion can be drawn from the Sylvester criterion, where the determinant of the Hessian is always negative,
\begin{equation}
\det\mathcal{H}=-\frac{1}{4s^3}< 0.
\end{equation}

\subsection{Local thermodynamic stability of Kerr black hole}

The admissible heat capacities of the Kerr black hole in ($s,j$) space are
\begin{align}\label{eqCMKerr}
C_m&=\tau \frac{\{s,m\}_{s,j}}{\{\tau,m\}_{s,j}} =\frac{s (s^2 -j^2)}{s^2 +j^2},
  \\\label{eqCOmKerr}
C_\omega& =\tau\frac{\{s,\omega \}_{s,j}}{\{\tau,\omega \}_{s,j}} =-\frac{2 s^3 (s^2 -j^2)}{(s^2 +j^2)^2},
   \\\label{eqCJKerr}
C_j&=\tau \frac{\{s,j\}_{s,j}}{\{\tau,j\}_{s,j}}=\frac{2s (s^2 -j^2) \left(s^2 +j^2\right)}{ 3j^4 +6j^2s^2 -s^4}.
\end{align}

For processes with constant mass $C_m$ is positive in the region of existence \eqref{eqRegExistKerr}, hence Kerr is locally stable above the blue line shown in Fig.\! \ref{KerrBlackHole}. Using the relation \eqref{eqMKerr}, together with the condition for existence \eqref{eqRegExistKerr}, we find $C_m$ as a function of one variable $s$ or $q$ and the constant mass $m_c$:
\begin{equation}
C_m= \frac{s(2s -m_c^2)}{m_c^2}= \frac{ m_c^4 -4j^2 +m_c^2\sqrt{m_c^4 -4j^2} }{2m_c^2}.
\end{equation}

For processes with constant angular velocity the heat capacity $C_{\omega}$ is always negative in the region of existence, thus the black hole is locally unstable for this kind of processes. Using \eqref{eqOKerr} we can find $C_\omega$ as a function of $s$ and the constant parameter $\omega_c$:
\begin{equation}
C_\omega=-2s \big( 1 -\omega_c^2s \big) \big( 1 -2 \omega_c^2s \big).
\end{equation}

Finally, local thermodynamic stability for processes with fixed angular momentum requires
\begin{equation}
|j|<s<|j| \sqrt{3+2\sqrt{3}}.
\end{equation}
This is the region between the blue and the orange lines (Fig.\! \ref{KerrBlackHole}), where Kerr is locally stable with respect to $C_j$. Above the orange line the black hole is locally unstable for such processes. The heat capacity $C_j(s)$ is a function of one variable $s$, because $j_c$ is a constant. 

\subsection{Critical curves and phase transitions of Kerr black hole}

The set of critical curves in ($s,j$) space are defined by
\begin{align}
C_m\to\left\{
\!\!\begin{array}{l}
 0,\,\,s\to |j|,\\
   s=m_c^2,\,\, j\to 0,
\end{array} 
\right. 
\,\,\, 
C_{\omega}\to\left\{
\!\!\begin{array}{l}
 0,\,\,s\to |j|,\\
   0,\,\, j\to 0,
\end{array} 
\right. 
\,\,\,
C_j\to\left\{
\!\!\!\begin{array}{l}
 \quad0,\,\,\,\,s\to |j|,\\
   \pm\infty,\,\, s\to  |j|\sqrt{3+2 \sqrt{3}} \mp 0.
\end{array} 
\right.
\end{align}

The heat capacities $C_m$, $C_{\omega}$ and $C_j$ change sign on the blue line $s=|j|$ (Fig.\! \ref{KerrBlackHole}), which corresponds to the extremal case. For process with constant mass, on the line $j= 0$ (the angular velocity is $\omega=0$) the Kerr black hole reduces to a new locally stable state with respect to $C_m$. The latter is not a Schwarzschild black hole, since the heat capacity $C_m$ differs from \eqref{HeatCapSch}. It is a regular limit and not a phase transition. 

For processes with constant angular velocity \eqref{eqOKerr}, the limit $j\to0$ leads to $s\to 0$, which is forbidden by the third law. 

Finally, for fixed angular momentum, the (orange) line $s=$ $|j|\sqrt{3+2 \sqrt{3}}$ is a Davies curve. It indicates a phase transition from locally stable to unstable thermodynamic state of the Kerr black hole with respect to $C_j$.

\section{Thermodynamic stability of Kerr-Newman solution}\label{secKN}

\subsection{Thermodynamics in Kerr-Newman spacetime}

The Kerr-Newman (KN) spacetime is the charged version of the Kerr solution with line element
\begin{align}
   \nonumber ds^2=&-\bigg(1-\frac{2 M r-Q^2}{\Sigma}\bigg) dt^2-\frac{(2 M r-Q^2) 2 a \sin^2\theta}{\Sigma} dt d\phi
    \\
    &+\frac{\Sigma}{\Delta} dr^2+\Sigma d\theta^2+\bigg(r^2+a^2+ \frac{(2 M r-Q^2)a^2\sin^2\theta}{\Sigma}\bigg) \sin^2 {\theta}d\phi^2,
\end{align}
where we have the notations:
\begin{equation}
    \Sigma=r^2+a^2 \cos^2\theta,\quad \Delta=r^2-2 M r+a^2+Q^2,\quad J=a M.
\end{equation}
In addition to the gravitational field, the KN black hole is surrounded by a stationary
electromagnetic field which is completely determined by the charge $Q$ and
the parameter $a$. The standard form of the KN thermodynamics is presented in Appendix \ref{appC}.
Here we prefer to work with  new set of parameters defined by
\begin{equation}\label{RescaledParKN}
m=2M, \quad \tau=2\pi T, \quad  s=\frac{S}{\pi}, \quad \omega=\Omega, \quad j=2J, \quad q=Q, \quad \phi=2\Phi,
\end{equation}
The latter renders the KN thermodynamics in the form $ dm=\tau ds +\omega dj +\phi dq$ with
\begin{align}
&m=\sqrt{\frac{j^2+\big(q^2+s\big)^2\,}{s}},
\\
&\tau=\frac{\partial m}{\partial s}\bigg|_{j,q}=\frac{s^2-\big(j^2+q^4\big)}{2\sqrt{ s^3\big(j^2 +(q^2+s)^2\big) }}, \\
&\omega=\frac{\partial m}{\partial j}\bigg|_{s,q}=\frac{j}{\sqrt{s\big(j^2+(q^2+s)^2\big)}}, \\
&\phi=\frac{\partial m}{\partial q}\bigg|_{s,j}=\frac{2q\big(q^2+s\big)}{\sqrt{s\big(j^2+(q^2+s)^2 \big)}}.
\end{align}

The region of existence of the Kerr-Newman black hole in $(s,j,q)$ space is now given by
\begin{equation}\label{eqKNExConS}
    s>\sqrt{j^2+q^4}.
\end{equation}
This condition is satisfied above the blue surface Fig. \ref{Cwq_Red}.

\subsection{Global thermodynamic instability of Kerr-Newman black hole}

The global thermodynamic stability is determined by the components of the  Hessian:
\begin{equation}
    \mathcal{H}(s,j,q) =\left(
\begin{array}{ccc}
\mathcal{H}_{ss} & \mathcal{H}_{sj} &\mathcal{H}_{sq} \\[5pt]
\mathcal{H}_{js}  & \mathcal{H}_{jj}  & \mathcal{H}_{jq} \\[5pt]
\mathcal{H}_{qs}  & \mathcal{H}_{qj}  & \mathcal{H}_{qq} 
\end{array}
\right)
=\left(
\begin{array}{ccc}
\frac{\partial^2 m}{\partial s^2} & \frac{\partial^2 m}{\partial s \partial j} & \frac{\partial^2 m}{\partial s\partial q} \\[5pt]
\frac{\partial^2 m}{\partial j\partial s} & \frac{\partial^2 m}{\partial j^2} & \frac{\partial^2 m}{\partial j \partial q} \\[5pt]
\frac{\partial^2 m}{\partial q\partial s} & \frac{\partial^2 m}{\partial q\partial j} & \frac{\partial^2 m}{ \partial q^2}
\end{array}
\right).
\end{equation}
 where the explicit expressions are given in Appendix \ref{appD}. The eigenvalues of the Hessian satisfy a cumbersome cubic equation,
\begin{align}
\nonumber {\lambda ^3} &- \big( {{\mathcal{H}_{jj}} + {\mathcal{H}_{qq}} + {\mathcal{H}_{ss}}} \big){\lambda ^2}
\\\nonumber
&+ \big( {{\mathcal{H}_{jj}}{\mathcal{H}_{qq}} + {\mathcal{H}_{jj}}{\mathcal{H}_{ss}} - \mathcal{H}_{jq}^2 + {\mathcal{H}_{qq}}{\mathcal{H}_{ss}} - \mathcal{H}_{sj}^2 - \mathcal{H}_{sq}^2} \big)\lambda 
\\
 &- {\mathcal{H}_{jj}}{H_{qq}}{\mathcal{H}_{ss}} + {\mathcal{H}_{jj}}\mathcal{H}_{sq}^2 - 2{\mathcal{H}_{jq}}{\mathcal{H}_{sj}}{\mathcal{H}_{sq}} + \mathcal{H}_{jq}^2{\mathcal{H}_{ss}} + {\mathcal{H}_{qq}}\mathcal{H}_{sj}^2 = 0,
\end{align}
which makes them difficult for an analytical treatment\footnote{Nevertheless, simple numerical study in the region of existence shows that they can differ by signs, thus confirming the result from the Sylvester criterion.}. Fortunately, we can use the conditions imposed by the  Sylvester criterion. In mass-energy representation global thermodynamic stability insists on positive definiteness of the Hessian of the mass. The latter suggests that the first level principal minors of the Hessian should satisfy
\begin{equation}
    \mathcal{H}_{ss}> 0,\quad  \mathcal{H}_{jj}> 0,\quad   \mathcal{H}_{qq}> 0,
\end{equation}
together with the conditions for the  second level principal minors:
\begin{align}\label{eqStrongerTDT}
\Delta_{s}=\left|\!
\begin{array}{cc}
\mathcal{H}_{jj} & \mathcal{H}_{jq}\\[5pt]
\mathcal{H}_{qj} & \mathcal{H}_{qq}\\
\end{array}
\!\right|> 0,
\quad
\Delta_{j}=\left|\!
\begin{array}{cc}
\mathcal{H}_{ss} & \mathcal{H}_{sq}\\[5pt]
\mathcal{H}_{qs} & \mathcal{H}_{qq}\\
\end{array}
\!\right|> 0,
\quad\Delta_{q}=\left|\!
\begin{array}{cc}
\mathcal{H}_{ss} & \mathcal{H}_{sj}\\[5pt]
\mathcal{H}_{js} & \mathcal{H}_{jj}\\
\end{array}
\!\right|> 0,
\end{align}
and the determinant of the Hessian itself:
\begin{equation}
\Delta=\det\mathcal{H}=\left|
\begin{array}{ccc}
\mathcal{H}_{ss} & \mathcal{H}_{sj} &\mathcal{H}_{sq} \\[5pt]
\mathcal{H}_{js}  & \mathcal{H}_{jj}  & \mathcal{H}_{jq} \\[5pt]
\mathcal{H}_{qs}  & \mathcal{H}_{qj}  & \mathcal{H}_{qq} 
\end{array}
\right|> 0.
\end{equation}
In this case, it only suffices to calculate the determinant of the Hessian and show that it is always negative in the region of existence (\ref{eqKNExConS}):
\begin{equation}
\det\mathcal{H}= -\,\frac{(s-q^2) (q^2+s)^2 + j^2 (3q^2 +s)}{2 \sqrt{s^7 \big(j^2 +(q^2+s)^2 \big)^3} }<0.
\end{equation}
Therefore the RN black hole is globally unstable from thermodynamic standpoint. 

\subsection{Local thermodynamic stability of Kerr-Newman black hole and critical points}

\begin{figure}[H]
\centering
\includegraphics[scale=0.35]{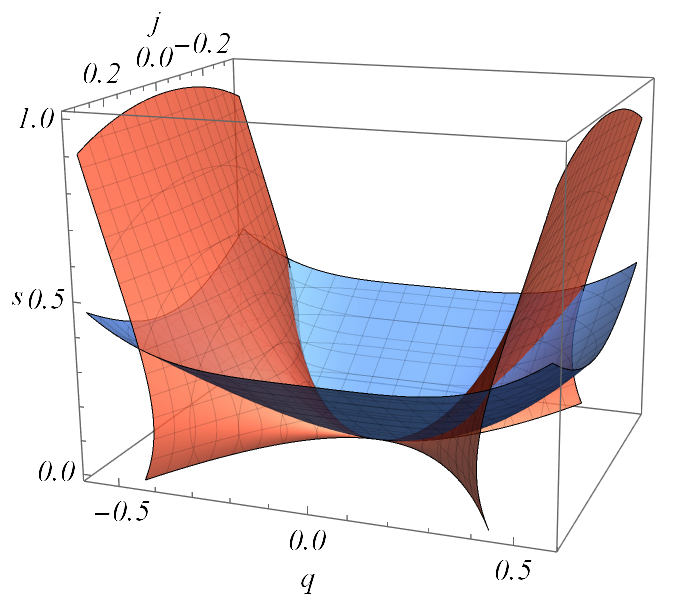}
\caption{ The region of existence of the KN black hole is above the blue surface $s=\sqrt{j^2+q^4}$. The red surface \eqref{Red_Surf} represents the Davies surface for $C_{\omega,q}$. One has local thermodynamic stability with respect to $C_{\omega,q}$ between the two surfaces. }
\label{Cwq_Red}
\end{figure}

\begin{figure}[H]
\centering
\includegraphics[scale=0.35]{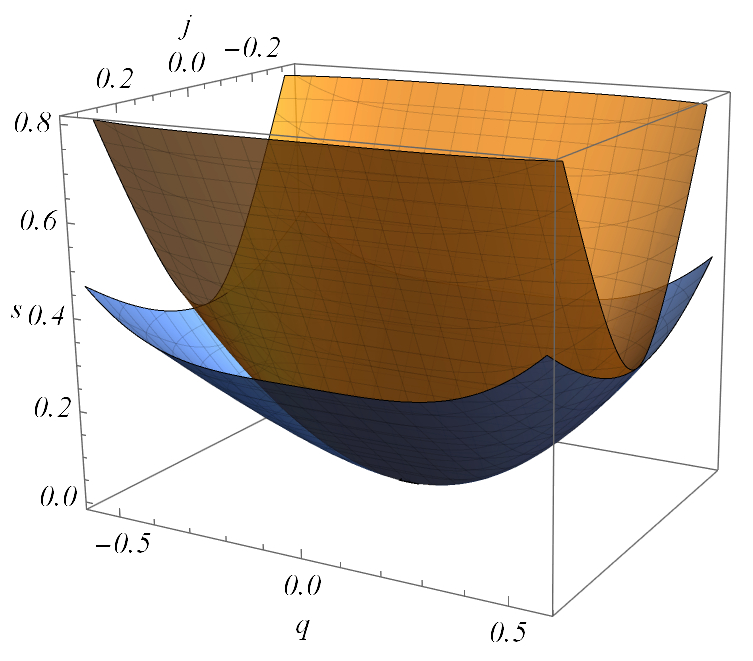}
\caption{ The orange surface \eqref{Orange_Surf} indicates the Davies surface for $C_{j,\phi}$. One has local thermodynamic stability with respect to $C_{j,\phi}$ between the blue and the orange surfaces. Above the orange surface the black hole is unstable. }
\label{Cjf_Orange}
\end{figure}

\begin{figure}[H]
\centering
\includegraphics[scale=0.35]{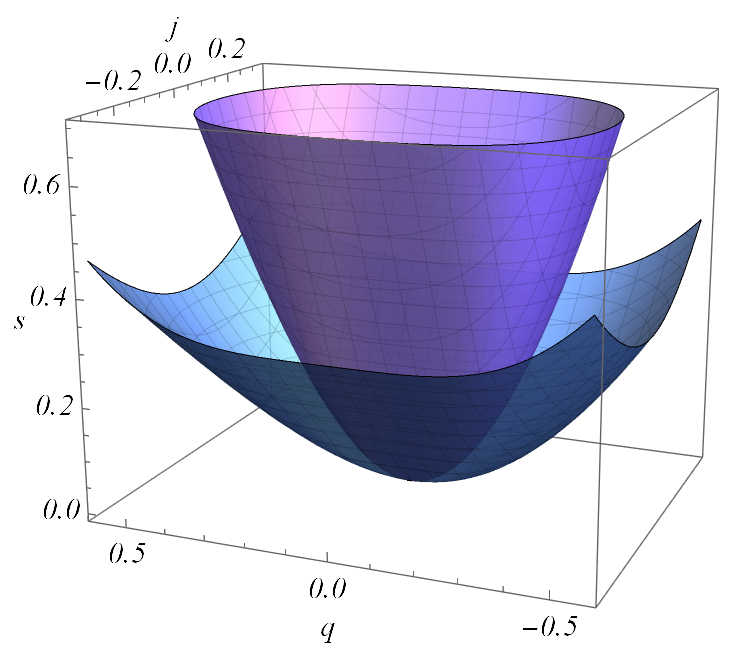}
\caption{The purple surface \eqref{Purple_Surf} represents the Davies surface for $C_{j,q}$. One has local thermodynamic stability with respect to $C_{j,q}$ between the blue and the purple surfaces. Above the purple surface the black hole is unstable. }
\label{Cjq_Purple}
\end{figure}

The regions of local thermodynamic stability for the Kerr-Newman black hole can be identified as the positive definiteness of  the admissible heat capacities in ($s,j,q$) space. There are a dozen of them due to the greater number of state quantities involved. In order to study them, we define the following surfaces in the ($s,j,q$) space:
\begin{itemize}
\item The blue surface (Figs.\ref{Cwq_Red}-\ref{Cjq_Purple}) defines the existence surface of the KN black hole:
\begin{equation}
    B(s,j,q)= s-\sqrt{j^2+q^4}=0.
\end{equation}
\item The red surface (Fig. \ref{Cwq_Red}) indicates the Davies surface for $C_{\omega, q}$:
\begin{equation}\label{Red_Surf}
R(s,j,q)= q^2(2s +3q^2) -s^2 -j^2=0.
\end{equation}
\item The orange surface (Fig. \ref{Cjf_Orange}) corresponds to the Davies surface for $C_{j,\phi}$:
\begin{equation}\label{Orange_Surf}
O(s,j,q)=2j^2 (q^2+s) (6q^2s +5q^4 +3s^2) +3j^4 (3 q^2+s) -(s-q^2) (q^2+s)^4=0.
\end{equation}
\item The purple surface (Fig. \ref{Cjq_Purple}). defines the Davies surface for $C_{j,q}$:
\begin{equation}\label{Purple_Surf}
P(s,j,q)=3\big(j^2+q^4\big)^2  +\big(8sq^2 +6s^2\big) \big(j^2+q^4\big) -s^4=0.
\end{equation}
\end{itemize}

The first set of four heat capacity of the KN black hole are related to processes with constant mass:
\begin{align}
&C_{m,\omega }=\tau \frac{\partial s}{\partial \tau}\bigg|_{m,\omega}=\tau \frac{\{s,m,\omega \}_{s,j,q}}{\{\tau,m,\omega \}_{s,j,q}}=\frac{s\big(q^2+s\big) \big(s^2-(j^2+q^4)\big)}{q^2\big(j^2+(q^2+s)^2\big)}, \\
&C_{m,j}=\tau \frac{\partial s}{\partial \tau}\bigg|_{m,j}=\tau \frac{\{s,m,j\}_{s,j,q}}{\{\tau,m,j\}_{s,j,q}}=\frac{s\big(q^2+s\big) \big(s^2-(j^2+q^4)\big)}{j^2(q^2+2s) +q^2(q^2+s)^2},\\
&C_{m,\phi }=\tau \frac{\partial s}{\partial \tau}\bigg|_{m,\phi}=\tau\frac{\{s,m,\phi \}_{s,j,q}}{\{\tau,m,\phi \}_{s,j,q}}=\frac{s\big(3q^2+s\big)\big(s^2-(j^2+q^4)\big)}{j^2\big(3q^2+s\big) +\big(q^2+s\big) \big(s^2+2sq^2+3q^4\big)}, \\
&C_{m,q}=\tau \frac{\partial s}{\partial \tau}\bigg|_{m,q}=\tau\frac{\{s,m,q\}_{s,j,q}}{\{\tau,m,q\}_{s,j,q}}=\frac{s\big(s^2-(j^2+q^4)\big)}{s^2+j^2+q^4}.
\end{align}
All of them  are positive in the region of existence \eqref{eqKNExConS}, which is above the blue surface depicted in Fig. \ref{Cwq_Red}. Therefore the KN system retains local thermodynamic stability with respect to fixed mass.  

The heat capacities with two constant conjugate parameters with respect to the charge and the rotation, are equal to each other, $C_{\omega,j}=C_{\phi,q}$:
\begin{align}
&C_{\omega,j}=\tau \frac{\partial s}{\partial \tau}\bigg|_{\omega,j}\!=\tau \frac{\{s,\omega ,j\}_{s,j,q}}{\{\tau,\omega ,j\}_{s,j,q}}=\frac{s\big(q^2+s\big) \big(s^2-(j^2+q^4)\big)}{\big(2q^2+s\big) \big(j^2+(q^2+s)^2\big)},  \\
&C_{\phi,q }=\tau \frac{\partial s}{\partial \tau}\bigg|_{\phi,q}=\tau\frac{\{s,\phi ,q\}_{s,j,q}}{\{\tau,\phi ,q\}_{s,j,q}}=\frac{s(q^2+s) \big(s^2-(j^2+q^4)\big)}{(2q^2+s)\big(j^2+(q^2+s)^2\big)}. 
\end{align}
They are also positive in the region of existence \eqref{eqKNExConS}, therefore the KN black hole is locally stable for such processes.

The heat capacity with respect to fixed angular velocity and electric potential
\begin{equation}
C_{\omega ,\phi}=\tau \frac{\partial s}{\partial \tau}\bigg|_{\omega,\phi}=\tau \frac{\{s,\omega ,\phi \}_{s,j,q}}{\{\tau,\omega ,\phi \}_{s,j,q}}=\frac{-2s \big(q^2+s\big)^3 \big(s^2-(j^2+q^4)\big)}{\big(j^2+(q^2+s)^2\big) \big(j^2(3q^2+s) +(s-q^2)(q^2+s)^2\big)}
\end{equation}
is always negative in the region of existence. Hence the black hole is locally unstable for a processes with constant angular velocity and electric potential. 

The heat capacity  with fixed angular velocity and charge
\begin{equation}
C_{\omega ,q}=\tau \frac{\partial s}{\partial \tau}\bigg|_{\omega,q}=\tau \frac{\{s,\omega ,q\}_{s,j,q}}{\{\tau,\omega ,q\}_{s,j,q}}=\frac{2s\big(q^2+s\big)^2 \big(s^2-(j^2+q^4)\big)}{ \big(j^2+(q^2+s)^2\big)R(s,j,q)}
\end{equation}
is positive between the blue and the red surfaces (Fig. \ref{Cwq_Red}), which is the region of local stability. Above these surfaces the KN black hole is unstable thermodynamically for such processes.

For a process with constant angular momentum and electric potential, the heat capacity 
\begin{equation}
C_{j,\phi }=\tau \frac{\partial s}{\partial \tau}\bigg|_{j,\phi}=\tau \frac{\{s,j,\phi \}_{s,j,q}}{\{\tau,j,\phi \}_{s,j,q}}
=\frac{2s\big(s^2-(j^2+q^4)\big) \big(j^2(3q^2+s) +(q^2+s)^3\big)}{O(s,j,q)}
\end{equation}
is positive between the blue and the orange surfaces (Fig. \ref{Cjf_Orange}), hence the black hole is locally stable in this region. Above the orange surfaces the black hole is unstable.

Finally, we consider a process with constant angular momentum and electric charge. The corresponding heat capacity
\begin{equation}
C_{j,q}=\tau \frac{\partial s}{\partial \tau}\bigg|_{j,q}=\tau \frac{\{s,j,q\}_{s,j,q}}{\{\tau,j,q\}_{s,j,q}}= \frac{2s\big(s^2-(j^2+q^4)\big) \big(j^2+(q^2+s)^2\big)}{P(s,j,q)}
\end{equation}
is positive between the blue and the purple surfaces (Fig. \ref{Cjq_Purple}), hence the KN black hole is locally stable against such fluctuations. Above the purple surface the black hole is unstable and can radiate.

\section{Conclusion}\label{secConcl}

We conducted a thorough investigation of the thermodynamic stability of standard black hole solutions within the framework of general relativity, employing established techniques from classical thermodynamics. 
While these methods are widely applied in the study of conventional systems, their utilization in gravitational contexts, particularly in the thermodynamics of black holes, remains relatively unexplored. Our approach offers a significantly more systematic and comprehensive analysis compared to the scattered results found in existing literature.

We employ two rigorous global criteria for equilibrium: the Hessian eigenvalue method and the Sylvester criterion, which evaluates the positive definiteness of the mass-energy Hessian quadratic form. We illustrate that, with the exception of the simplest case of the Schwarzschild black hole, which is consistently thermodynamically unstable, Reissner-Nordstr\"om, Kerr, and Kerr-Newman solutions exhibit local stability in certain subregions against fluctuations concerning specific fixed parameters. However, comprehensive assessments of global stability reveal that all solutions in general relativity are globally unstable from a classical perspective. This situation does not appear to improve upon considering additional thermodynamic parameters in the Kerr-Newman black hole solution. Thus, it can be inferred that Schwarzschild, Kerr, and Reissner-Nordstr\"om black holes inherit their global thermodynamic instability from the Kerr-Newman black hole.

Our methodology offers the benefit of addressing perils often overlooked, which could lead to incomplete or incorrect conclusions regarding system stability. Firstly, we show that semi-definite criteria could lead to certain  contradictions with the stability of the system and should be considered with care. In essence, they allow the system to settle in the neighborhood of some of the classically forbidden regions in the state space, as shown for the RN black hole in Section \ref{RNsect}. Furthermore, we highlight potential inconsistencies arising from the consideration of partial stability conditions, as demonstrated again in Section \ref{RNsect}.

 Moreover, we emphasize the significance of evaluating all admissible heat capacities, as they provide crucial information about phase transition points and the black hole's responses to various perturbations. However, as demonstrated by \cite{Avramov:2023wyi}, the positivity of all heat capacities does not guarantee global stability of the black hole. This prompts us to suggest that solely examining heat capacities may not suffice for studying the thermodynamic stability of black holes. It is plausible that a comprehensive assessment involving all thermodynamic response functions, such as latent heats and compressibilities is necessary. We plan to investigate this matter further in a separate study.

This method has recently been employed to investigate the AdS family of black hole solutions, wherein the cosmological constant emerges naturally \cite{Avramov:2024hys}. In the aforementioned study, the authors underscore the crucial role played by the cosmological constant in stabilizing the thermodynamics of these black holes. A logical step forward from this point is to regard the cosmological constant as a thermodynamic variable, thereby introducing an effective pressure. We will explore this framework in the third installment of this series of papers. Indeed, our methodology lends itself readily to the extension of such cases, wherein the complexity of the thermodynamic state space is further heightened.

This paper marks the initial step in a series of planned investigations into the thermodynamic stability of black holes within modified theories of gravity and holography. In addition it would be intriguing to extend the full generic criteria to explore other types of black hole solutions, including,  but not limited to quantum-corrected black holes,   hairy black holes, black holes in lower and higher dimensions, regular black holes, black holes with extended thermodynamics, and related systems.

Furthermore, within the framework of holography, the dual quantum field theory can naturally undergo finite temperature embedding, induced by the corresponding black hole in the bulk. A prime example is the Kerr-AdS/QGP correspondence \cite{Casalderrey-Solana:2011dxg,DeWolfe:2013cua, Arefeva:2016rob, Dolan:2016ech, Golubtsova:2022ldm}, where the characteristics of strongly correlated quark gluon plasma can be explored within the supergravity approximation. It would be intriguing to examine the thermodynamic stability of such systems, offering potential insights into nonperturbative quantum effects beyond the constraints of the supergravity approximation.

Another avenue of investigation   involves examining the interplay between thermodynamic and dynamic stability. Investigating fluctuation theory and its connection to thermodynamic information geometry could provide valuable insights as well. Finally, one could also study the relationship between thermodynamic stability and the holographic complexity of black holes.

\section*{Acknowledgments}

The authors would like to thank Goran Djordjevic, Dragoljub Dimitrijevic and the SEENET-MTP for the warm hospitality during BPU11 and satellite events. We are very grateful to S. Yazadjiev, D. Marvakov, V. Popov, P. Ivanov, G. Gyulchev, P. Nedkova, K. Staykov and I. Iliev for their useful comments and discussions. H. D. thankfully acknowledges the support by the NSF grant H28/5 and the support by the program “JINR-Bulgaria” of the Bulgarian Nuclear Regulatory Agency. V. A. was partially supported by Sofia University grant 80-10-150 and the SEENET-MTP - ICTP Program NT03. V. A. also gratefully acknowledges the support by the Simons Foundation and the International Center for Mathematical Sciences in Sofia for the various annual scientific events. M. R., R. R. and T. V. were fully financed  by the European Union- NextGeneration EU, through the National Recovery and Resilience Plan of the Republic of Bulgaria, project BG-RRP-2.004-0008-C01.

\appendix
\section{Sylvester's criterion for systems with three independent parameters. Local vs. global thermodynamic stability}\label{appA}

Following Callen \cite{callen2006thermodynamics} we define the sufficient strong intrinsic global condition for thermodynamic stability as the strict  convexity/concavity
of the energy/entropy:
\begin{align}\label{eqIntrGlobalE}
&E\!\big(\!E^1 + \Delta E^1,\!E^2 + \Delta E^2\!,...\big) + E\!\big(\!E^1 - \Delta E^1,\!E^2- \Delta E^2\!,...\big)> 2E (\!E^1, E^2, ...),
\\\label{eqIntrGlobalS}
&S\big(S^1 + \Delta S^1,S^2 + \Delta S^2,...\big) + S\big(S^1 - \Delta S^1,S^2- \Delta S^2, ...\big)< 2S (S^1, S^2,...),
\end{align}
which should be valid for all admissible values of the parameters and fluctuations $\Delta E^a=E^a-\bar E^a$ or $\Delta S^a=S^a-\bar S^a$. The sufficient  (differential) conditions  for thermodynamic stability follow from (\ref{eqIntrGlobalE}) and (\ref{eqIntrGlobalS}) by Taylor expansion up to second order in the fluctuations $\Delta E^a$ (or $\Delta S^a$) and are given by the Sylvester criterion for positive/negative definiteness of the Hessians of the energy/entropy. This criterion is in general less restrictive than the convexity/concavity conditions above, but it is sufficient to assure strict convexity/concavity of the energy/entropy in certain intervals of the independent parameters. For this reason we call Sylvester criterion ``global'', because it establishes strict convexity/concavity and hence global equilibrium of the system only within these intervals. 

On the other hand, due to the fact that the components of the corresponding Hessians can be related to the thermodynamic coefficients such as heat capacities and compresibilities, we can define local equilibrium only by considering positiveness of these coefficients in any given ensemble. Strictly (global) convex/concave functions can be achieved in a certain parameter region if and only if all thermodynamic coefficients are stable (positive). Therefore we make a bit artificial distinction between local and global thermodynamic stability, due to the fact that one expects the Sylvester criterion to fail when one or more of the thermodynamic coefficients are negative\footnote{Nevertheless, there are examples of black hole systems, where all heat capacities of the system are positive, but one does not have global stability. We will show this in a subsequent paper in this line of investigations.}. 

In energy representation the energy defines a global convex function, thus the Hessian of the energy is positive definite quadratic form. In this case Sylvester's criterion states that all  the principal minors $\Delta_k>0$ of the Hessian of the energy must be strictly positive. In entropy representation this criterion has alternating signs $(-1)^k \Delta_k>0$ due to the fact that entropy is globally concave function.

Let us show what this implies for $n=3$ parametric thermodynamics in the energy representation. In this case the energy $E$ is a function of  its natural parameters $(E^1, E^2,E^3)$ and its  Hessian $ \mathcal{H}$ is the following $3\times 3$ symmetric matrix
\begin{equation}
\mathcal{H}(\vec E)=\left(
	\begin{array}{ccc}
		\frac{\partial^2 E}{(\partial E^1)^2}\big|_{E^2,E^3} & \frac{\partial^2 E}{\partial E^1 \partial E^2}\big|_{E^3} &\frac{\partial^2 E}{\partial E^1 \partial E^3}\big|_{E^2}\\[5pt]
		\frac{\partial^2 E}{\partial E^1 \partial E^2}\big|_{E^3} & \frac{\partial^2 E}{(\partial E^2)^2}\big|_{E^1,E^3} & \frac{\partial^2 E}{\partial E^2 \partial E^3}\big|_{E^1}\\[5pt]
	\frac{\partial^2 E}{\partial E^1 \partial E^3}\big|_{E^2} & \frac{\partial^2 E}{\partial E^2 \partial E^3}\big|_{E^1} & \frac{\partial^2 E}{(\partial E^3)^2}\big|_{E^1,E^2}
	\end{array}
	\right).
\end{equation}
According to the Sylvester criterion the fist condition for global thermodynamic stability requires the first level principal minors of $\hat{\mathcal{H}}$ be strictly positive:
\begin{equation}\label{eqGlobalCondE}
\mathcal H_{11}=\frac{\partial^2 E}{(\partial E^1)^2}\bigg|_{E^2,E^3}>0,
\quad \mathcal H_{22}=\frac{\partial^2 E}{(\partial E^2)^2}\bigg|_{E^1, E^3}>0,
\quad 	\mathcal H_{33}=\frac{\partial^2 E}{(\partial E^3)^2}\bigg|_{E^1, E^2}>0.
\end{equation}
One should consider the previous conditions, together with the restrictions on the determinants of second level principal minors:
\begin{align}\label{eqStrongerTD}
&\Delta_3=\left|
	\begin{array}{cc}
		\frac{\partial^2 E}{(\partial E^1)^2}\big|_{E^2,E^3} & \frac{\partial^2 E}{\partial E^1 \partial E^2}\big|_{E^3} \\[5pt]
		\frac{\partial^2 E}{\partial E^1 \partial E^2}\big|_{E^3} & \frac{\partial^2 E}{(\partial E^2)^2}\big|_{E^1,E^3}\\
	\end{array}
	\right|> 0, 
\\
&\Delta_2=\left|
\begin{array}{cc}
	\frac{\partial^2 E}{(\partial E^1)^2}\big|_{E^2,E^3} & \frac{\partial^2 E}{\partial E^1 \partial E^3}\big|_{E^2} \\[5pt]
	\frac{\partial^2 E}{\partial E^1 \partial E^3}\big|_{E^2} & \frac{\partial^2 E}{(\partial E^3)^2}\big|_{E^1,E^2}\\
\end{array}
\right|> 0, 
\\
&\Delta_1=\left|
\begin{array}{cc}
	\frac{\partial^2 E}{(\partial E^2)^2}\big|_{E^1,E^3} & \frac{\partial^2 E}{\partial E^2 \partial E^3}\big|_{E^1} \\[5pt]
	\frac{\partial^2 E}{\partial E^2 \partial E^3}\big|_{E^1} & \frac{\partial^2 E}{(\partial E^3)^2}\big|_{E^2,E^3}\\
\end{array}
\right|> 0.
\end{align}
Here, the lower index in $\Delta_{i}$ indicates that the $i^{th}$ row and column of the Hessian have been removed.  The final part of the Sylvester criterion is a condition on the determinant of the Hessian itself:
\begin{equation}\label{eqGlobalCondE1}
\Delta=\det\mathcal{H}=\left|
	\begin{array}{ccc}
		\frac{\partial^2 E}{(\partial E^1)^2}\big|_{E^2,E^3} & \frac{\partial^2 E}{\partial E^1 \partial E^2}\big|_{E^3} &\frac{\partial^2 E}{\partial E^1 \partial E^3}\big|_{E^2}\\[5pt]
		\frac{\partial^2 E}{\partial E^1 \partial E^2}\big|_{E^3} & \frac{\partial^2 E}{(\partial E^2)^2}\big|_{E^1,E^3} & \frac{\partial^2 E}{\partial E^2 \partial E^3}\big|_{E^1}\\[5pt]
		\frac{\partial^2 E}{\partial E^1 \partial E^3}\big|_{E^2} & \frac{\partial^2 E}{\partial E^2 \partial E^3}\big|_{E^1} & \frac{\partial^2 E}{(\partial E^3)^2}\big|_{E^1,E^2}
	\end{array}
	\right|>0.
\end{equation}

Similar conditions can be stated for $n=3$ in the entropy representation. In this case the first part of the Sylvester criterion yields
\begin{align}\label{eqGlobalCondS}
\mathcal H^{S}_{11}=\frac{\partial^2 S}{(\partial S^1)^2}\bigg|_{S^2,S^3}<0,
\quad \mathcal H^{S}_{22}=\frac{\partial^2 S}{(\partial S^2)^2}\bigg|_{S^1, S^3}<0,
\quad 	\mathcal H^{S}_{33}=\frac{\partial^2 S}{(\partial S^3)^2}\bigg|_{S^1, S^2}<0,
\end{align}
which just reflects the fact that entropy is a concave function along its natural parameters $(S^1, S^2, S^3)$. The second part of the criterion
requires
\begin{align}\label{eqStrongerTDS}
&\Delta^S_3=\left|
\begin{array}{cc}
	\frac{\partial^2 S}{(\partial S^1)^2}\big|_{S^2,S^3} & \frac{\partial^2 S}{\partial S^1 \partial S^2}\big|_{S^3} \\[5pt]
	\frac{\partial^2 S}{\partial S^1 \partial S^2}\big|_{S^3} & \frac{\partial^2 S}{(\partial S^2)^2}\big|_{S^1,S^3}\\
	\end{array}
	\right|> 0, 
\\
&\Delta^S_2=\left|
\begin{array}{cc}
	\frac{\partial^2 S}{(\partial S^1)^2}\big|_{S^2,S^3} & \frac{\partial^2 S}{\partial S^1 \partial S^3}\big|_{S^2} \\[5pt]
	\frac{\partial^2 S}{\partial S^1 \partial S^3}\big|_{S^2} & \frac{\partial^2 S}{(\partial S^3)^2}\big|_{S^1,S^2}\\
\end{array}
\right|> 0, 
\\
&\Delta^S_1=\left|
\begin{array}{cc}
	\frac{\partial^2 S}{(\partial S^2)^2}\big|_{S^1,S^3} & \frac{\partial^2 S}{\partial S^2 \partial S^3}\big|_{S^1} \\[5pt]
	\frac{\partial^2 S}{\partial S^2 \partial S^3}\big|_{S^1} & \frac{\partial^2 S}{(\partial S^3)^2}\big|_{S^2,S^3}\\
\end{array}
\right|> 0.
\end{align}
Finally, the third part is
\begin{equation}\label{eqGlobalCondS1}
\Delta^{S}=\det\mathcal{H}^S=\left|
	\begin{array}{ccc}
		\frac{\partial^2 S}{(\partial S^1)^2}\big|_{S^2,S^3} & \frac{\partial^2 S}{\partial S^1 \partial S^2}\big|_{S^3} &\frac{\partial^2 S}{\partial S^1 \partial S^3}\big|_{S^2}\\[5pt]
		\frac{\partial^2 S}{\partial S^1 \partial S^2}\big|_{S^3} & \frac{\partial^2 S}{(\partial S^2)^2}\big|_{S^1,S^3} & \frac{\partial^2 S}{\partial S^2 \partial S^3}\big|_{S^1}\\[5pt]
		\frac{\partial^2 S}{\partial S^1 \partial S^3}\big|_{S^2} & \frac{\partial^2 S}{\partial S^2 \partial S^3}\big|_{S^1} & \frac{\partial^2 S}{(\partial S^3)^2}\big|_{S^1,S^2}
	\end{array}
	\right|<0.
\end{equation}

Note the differences in the signs of (\ref{eqGlobalCondE}) and (\ref{eqGlobalCondS}), and also between (\ref{eqGlobalCondE1}) and (\ref{eqGlobalCondS1}), which are due to the  convex/concave nature of the energy and the entropy.


\section{Nambu brackets}\label{appB}

The Nambu brackets generalizes the Poisson brackets for three or more variables. In general they account for the determinant of the Jacobian when working in certain coordinates, i.e.

\begin{equation}
\{f,x^1,...,x^{ n-1}\}_{y^1,y^2,...,y^n}=\left|
\begin{array}{cccc}
	\frac{\partial f}{\partial y^1}\big|_{y^2,y^3,...,y^n}& \frac{\partial f}{\partial y^2}\big|_{{y^1,y^3,...,y^n}} & \cdots  & \frac{\partial f}{\partial y^n}\big|_{{y^1,y^2,..., y^{n-1}}} \\[5pt]
	\frac{\partial x^1}{\partial y^1}\big|_{{y^2,y^3,...,y^n}}	& 	\frac{\partial x^1}{\partial y^2}\big|_{{y^1,y^3,...,y^n}} &  \cdots & \frac{\partial x^1}{\partial y^n}\big|_{{y^1,y^2,...,y^{n-1}}} \\[5pt]
	\vdots	& \vdots  &   & \vdots\\[5pt]
	\frac{\partial x^{ n-1}}{\partial y^1}\big|_{{y^2,y^3,...,y^n}}	&  	\frac{\partial x^{ n-1}}{\partial y^2}\big|_{{y^1,y^3,...,y^n}}&  \cdots & 	\frac{\partial x^{ n-1}}{\partial y^n}\big|_{{y^1,y^2,...,y^{n-1}}}
	\end{array}
	\right|.
\end{equation}
For example, for $n=2$ one has
\begin{equation}
    \{f,x\}_{u,v}=   \left|
\begin{array}{cc}
\frac{\partial f}{\partial u}\big|_v& \frac{\partial f}{\partial v}\big|_u\\[5pt]
\frac{\partial x}{\partial u}\big|_v & \frac{\partial x}{\partial v}\big|_u\end{array}\right|
=\frac{\partial f}{\partial u}\bigg|_v 
 \frac{\partial x}{\partial v}\bigg|_u-\frac{\partial f}{\partial v}\bigg|_u \frac{\partial x}{\partial u}\bigg|_v.
\end{equation}

For $n=3$ one finds:
\begin{equation}
 \{f,x,y\}_{u,v,w}=\left|
	\begin{array}{ccc}
\frac{\partial f}{\partial u}\big|_{v,w} & \frac{\partial f}{\partial v}\big|_{u,w}  &\frac{\partial f}{\partial w}\big|_{u,v} \\[5pt]
\frac{\partial x}{\partial u}\big|_{v,w}  & \frac{\partial x}{\partial v}\big|_{u,w}  & \frac{\partial x}{\partial w}\big|_{u,v} \\[5pt]
\frac{\partial y}{\partial u}\big|_{v,w}  & \frac{\partial y}{\partial v}\big|_{u,w} & \frac{\partial y}{\partial w}\big|_{u,v} 
\end{array}
	\right|.
 \end{equation}
%
%

\section{Standard KN thermodynamics and existence conditions}\label{appC}

The existence of the event horizon of KN black hole at
\begin{equation}
r_+=M+\sqrt{M^2-Q^2-a^2}=\frac{1}{M}\big(M^2+\sqrt{M^4-Q^2 M^2-J^2}\big)>0
\end{equation}
imposes several constraints on the parameters of the black hole. Assuming $M>0$ with respect to the angular momentum $J$ one has
\begin{equation}\label{eqKNexistsCond1}
 -M\sqrt{M^2-Q^2}<J<M\sqrt{M^2-Q^2}\quad \text{and}\quad  M>|Q|  .   
 \end{equation}
With respect to the charge $Q$ one has
\begin{equation}\label{eqKNexistsCond2}
-\frac{\sqrt{M^4-{J^2}}}{M}<Q< \frac{\sqrt{M^4-{J^2}}}{M}\quad \text{and}\quad M>\sqrt{| J| }.
\end{equation}
Finally, with respect to the mass $M$ one has
\begin{equation}\label{eqKNexistsCond3}
   M>\frac{\sqrt{\sqrt{4 J^2+Q^4}+Q^2}}{\sqrt{2}}.
\end{equation}

In energy representation the mass $M$ is a function of $(S,J,Q)$, 
\begin{equation}
M=\sqrt{\frac{{4 \pi ^2 J^2+\left(\pi  Q^2+S\right)^2}}{{4\pi S }}}.
\end{equation}
Consequently, one can write the equations of state for $T$, $\Omega$ and $\Phi$ as
\begin{align}
& T=\frac{\partial M}{\partial S}\bigg|_{J,Q}=\frac{S^2-\pi ^2 \left(4 J^2+Q^4\right)}{4 \sqrt{\pi } S^{3/2} \sqrt{4 \pi ^2 J^2+\left(\pi  Q^2+S\right)^2}},
\\
&\Omega=\frac{\partial M}{\partial J}\bigg|_{S,Q}=\frac{2 \pi ^{3/2} J}{\sqrt{S \left(4 \pi ^2 J^2+\left(\pi  Q^2+S\right)^2\right)}},
\\
& \Phi=\frac{\partial M}{\partial Q}\bigg|_{S,J}=\frac{\sqrt{\pi } Q \left(\pi  Q^2+S\right)}{\sqrt{S \big(4\pi^2J^2 + (\pi Q^2 +S)^2 \big)}},
\end{align}
with first law of thermodynamics given by
\begin{equation}\label{mass1}
dM=T dS+\Omega dJ+\Phi dQ.
\end{equation}
\section{The Hessian of the mass for KN black hole}\label{appD}

The Hessian of the mass for the Kerr-Newman solution takes the following form
\begin{equation}
    \mathcal{H}(s,j,q) =\left(
\begin{array}{ccc}
\mathcal{H}_{ss} & \mathcal{H}_{sj} &\mathcal{H}_{sq} \\[5pt]
\mathcal{H}_{js}  & \mathcal{H}_{jj}  & \mathcal{H}_{jq} \\[5pt]
\mathcal{H}_{qs}  & \mathcal{H}_{qj}  & \mathcal{H}_{qq} 
\end{array}
\right)
=\left(
\begin{array}{ccc}
\frac{\partial^2 m}{\partial s^2} & \frac{\partial^2 m}{\partial s \partial j} & \frac{\partial^2 m}{\partial s\partial q} \\[5pt]
\frac{\partial^2 m}{\partial j\partial s} & \frac{\partial^2 m}{\partial j^2} & \frac{\partial^2 m}{\partial j \partial q} \\[5pt]
\frac{\partial^2 m}{\partial q\partial s} & \frac{\partial^2 m}{\partial q\partial j} & \frac{\partial^2 m}{ \partial q^2}
\end{array}
\right).
\end{equation}
 where the explicit expressions for the components are given by
\begin{align}\label{eqHssKN}
&\mathcal{H}_{ss} =\frac{3\big(j^2+q^4\big)^2  +s\big(8q^2 +6s\big) \big(j^2+q^4\big) -s^4}{4\sqrt{s^5 \big(j^2 +(q^2+s)^2\big)^3}},
\\\label{eqHjjKN}
&\mathcal{H}_{jj}=\frac{\big(q^2+s\big)^2}{\sqrt{s\big(j^2 +(q^2+s)^2\big)^3}},
\\\label{eqHqqKN}
&\mathcal{H}_{qq}=2\,\frac{\big(q^2+s\big)^3 +j^2\big(3q^2+s\big)}{\sqrt{s\big(j^2+(q^2+s)^2\big)^3}},
\\
&\mathcal{H}_{sj}=\mathcal{H}_{js}=-\,j\,\frac{ j^2 +(q^2+s)(q^2+3s) }{2\sqrt{s^3\big(j^2 +(q^2+s)^2\big)^3}},
\\
&\mathcal{H}_{sq}=\mathcal{H}_{qs}= -\,q\, \frac{ (q^2+s)^3 -j^2(s-q^2) }{\sqrt{s^3\big(j^2+(q^2+s)^2\big)^3}},
\\
&\mathcal{H}_{jq}=\mathcal{H}_{qj}=-\, \frac{2jq\big(q^2+s\big)}{\sqrt{s\big(j^2+(q^2+s)^2\big)^3}}.
\end{align}

The expressions of second level minors are
\begin{align}
&\Delta_s=\frac{2 \big(q^2 +s\big)^3}{s \big(j^2 +(q^2+s)^2 \big)^2}, \\
&\Delta_j=\frac{2j^2 (q^2+s) (6q^2s +5q^4 +3s^2) +3j^4 (3 q^2+s) -(s-q^2) (q^2+s)^4}{2 s^3 \big(j^2 +(q^2+s)^2 \big)^2}, \\
&\Delta_q=\frac{ q^2(2s +3q^2) -s^2 -j^2}{4s^3 \big(j^2 +(q^2+s)^2 \big)}.
\end{align}
%


\providecommand{\href}[2]{#2}\begingroup\raggedright\endgroup

\end{document}